\documentclass[pre,superscriptaddress,twocolumn,amsmath,amssymb,floatfix,showpacs]{revtex4}

\usepackage{graphicx}
\usepackage{dcolumn}
\usepackage{bm}
\usepackage[paperwidth=210mm, paperheight=297mm, centering, hmargin=2cm, vmargin=2cm]{geometry}
\begin{document}

\title{Impact of Impulse Stops on Pedestrian Flow} 
\author{Jaeyoung Kwak}
\email{jaeyoung.kwak@aalto.fi}
\affiliation{Department of Civil and Environmental Engineering, Aalto University, Espoo, Finland}
\author{Hang-Hyun Jo}
\affiliation{BK21plus Physics Division and Department of Physics, Pohang University of Science and Technology, Pohang 37673, Republic of Korea}
\affiliation{Department of Computer Science, Aalto University School of Science, P.O. Box 15400, Finland}
\author{Tapio Luttinen}
\affiliation{Department of Civil and Environmental Engineering, Aalto University, Espoo, Finland}
\author{Iisakki Kosonen}
\affiliation{Department of Civil and Environmental Engineering, Aalto University, Espoo, Finland}

\date{November 15, 2015}

\begin{abstract}
We numerically study the impact of impulse stops on pedestrian flow for a straight corridor with multiple attractions. The impulse stop is simulated by the switching behavior model, a function of the social influence strength and the number of attendees near the attraction. When the pedestrian influx is low, one can observe a stable flow where attendees make a complete stop at an attraction and then leave the attraction after a certain amount of time. When the pedestrian influx is high, an unstable flow is observed for strong social influence. In the unstable flow, attendees near the attraction are crowded out from the clusters by others due to the interpersonal repulsion. The expelled pedestrians impede the pedestrian traffic between the left and right boundaries of the corridor. These collective patterns of pedestrian flow are summarized in a schematic phase diagram.
\end{abstract}


\maketitle

\section{Introduction}
\label{sec:1}
Walking is a fundamental activity of human life, not only for moving between places but also in interactions with surrounding environments. While walking to destinations, pedestrians may be influenced by attractive stimuli such as artworks and shop displays. Some pedestrians may shift their attention to such attractions, opting to stop walking and making an impulse stop to join an attraction~\cite{Borgers_1986}.

According to previous studies~\cite{Gallup_2012, Milgram_1969}, it has been reported that a growing number of attendees around an attraction are likely to attract more passersby to the attraction, inferring that impulse stopping pedestrians can be affected by others’ choice. It has been widely accepted that having more store visitors likely attracts more passerby to the store in that a growing number of visitors increases the possibility of passersby visiting the store. Therefore, marketing strategies have focused on increasing the number of the impulse stopping visitors~\cite{Bearden_1989}.

By means of numerical simulations, we have investigated the impact of impulse stops on pedestrian flow for a straight corridor with multiple attractions. This study employs the switching behavior model, as shown in the next section. In Sect.~\ref{sec:3}, we analyze the spatial distribution of the pedestrian flow, characterize the collective patterns, and then summarize the results with a schematic phase diagram. Finally, we discuss the findings of this study in the section following the results (see Sect.~\ref{sec:4}).

\section{Model}
\label{sec:2}
\subsection{Switching Behavior}
\label{subsec:1}
Similar to the sigmoidal choice rule~\cite{Gallup_2012, Milgram_1969}, the probability of joining an attraction $P_{a}$ is a function of the number of pedestrians who have already joined $N_{a}$ and the number of pedestrians not stopping by the attraction $N_{0}$:

\begin{eqnarray}\label{eq:P_a}
P_{a} = \frac{s(N_{a}+K_a)}{({N_{0}+K_0})+s({N_{a}+K_a})}.
\end{eqnarray}

This suggests that larger $N_{a}$ likely yields higher joining probability. In order to prevent indeterminate cases with $N_{a} = N_{0} = 0$, two baseline values $K_a$ and $K_0$ are introduced for $N_{a}$ and $N_{0}$. Here $s > 0$ is the strength of the social influence that can be also understood as pedestrians' awareness of the attraction. According to previous studies~\cite{Gallup_2012, Kaltcheva_2006, Milgram_1969}, we assumed that the strength of social influence can be different for different situations and can be controlled in the presented model. After joining the attraction, the individual will then stay near the attraction for an exponentially distributed time with an average of $t_{d}$, similar to previous works~\cite{Gallup_2012, Helbing_PRE1995, Kwak_PLOS2015}. 

\subsection{Pedestrian Movement}
\label{subsec:2}
According to the social force model~\cite{Helbing_PRE1995}, the velocity $\vec{v}_i(t)$ of pedestrian $i$ at time $t$ is given by the following equation:
\begin{eqnarray}\label{eq:EoM}
\frac{\mathrm{d} \vec{v}_i(t)}{\mathrm{d} t} = \frac{v_d\vec{e}_{i}-\vec{v}_{i}(t)}{\tau} +\sum_{j\neq i}^{ }{\vec{f}_{ij}}+\sum_{B}^{ }{\vec{f}_{iB}}.
\end{eqnarray}

Here the first term on the right-hand side indicates the driving force describing the tendency of pedestrian $i$ moving toward his destination with the desired speed ${v}_{d}$ and an unit vector $\vec{e}_{i}$ pointing to the desired direction. The relaxation time $\tau$ controls how fast pedestrian $i$ adapts its velocity to the desired velocity. The repulsive force terms $\vec{f}_{ij}$ and $\vec{f}_{iB}$ reflect his tendency to keep certain distance from other pedestrian $j$ and the boundary $B$, e.g., wall and obstacles. A more detailed description of the pedestrian movement model can be found in previous works~\cite{Helbing_PRE1995, Johansson_2007, Kwak_PRE2013, Kwak_PLOS2015}.

\subsection{Numerical simulation setup}
\label{subsec:3}
Each pedestrian is modeled by a circle with radius $r_i = 0.25$~m. Pedestrians move in a corridor of length 55~m and width 6~m in the horizontal direction. They move with desired speed $v_d = 1.2$~m/s and with relaxation time $\tau = 0.5$~s, and their speed cannot exceed $v_{\rm max} = 2.0$~m/s. The desired direction points from the left to the right boundary of the corridor for one half of population and the opposite direction for the other half. On the lower wall of the corridor, three attractions are placed for every 15~m. The number of pedestrians in the corridor is associated with the pedestrian influx $q$, i.e., the arrival rate of pedestrians entering the corridor. The pedestrian arrival rate is assumed to follow a shifted exponential distribution $h = 1/q$ with a minimum headway $h_0 = 0.5$~s per unit width based on previous works~\cite{Luttinen_1999, May_1990}. 

\begin{figure}[!t]
\includegraphics[width=.7\columnwidth]{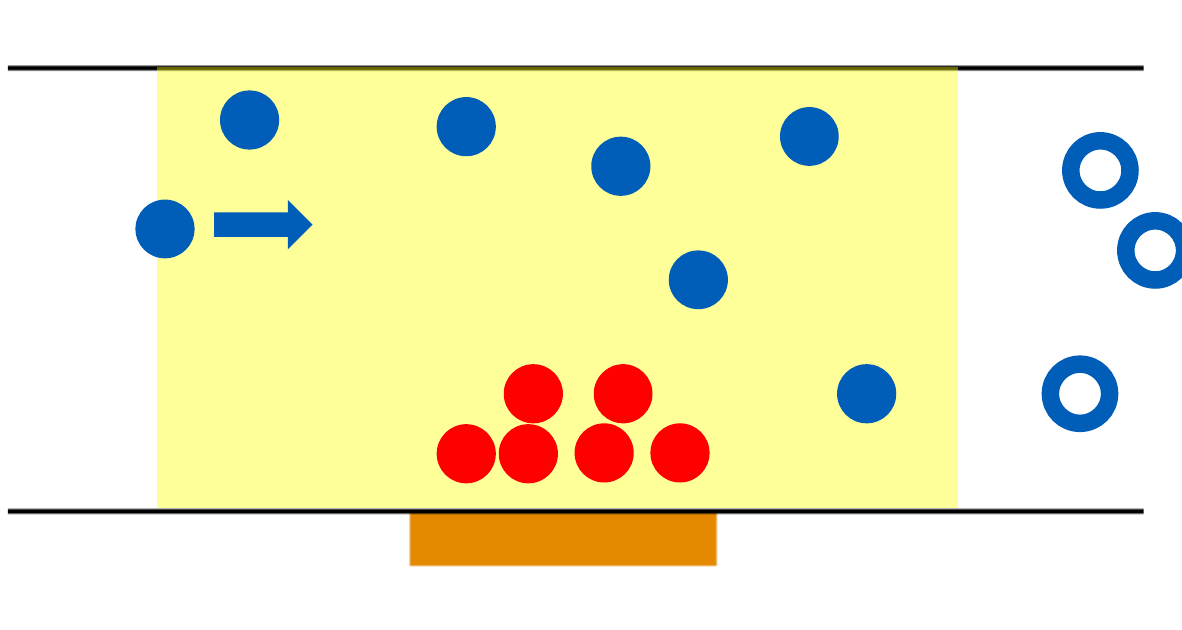}
\caption{(Color online) A schematic representation of the area of influence.}
\label{fig:influenceArea} 
\end{figure}

The joining probability (Eq.~\ref{eq:P_a}) is updated with the social force model (Eq.~\ref{eq:EoM}) for each simulation time step of 0.05~s. The individual can decide whether he will join the attraction when he enters the area of influence (see Fig.~\ref{fig:influenceArea}). The area of influence is defined as a square area of 15~m by 6~m, and its horizontal center coincides with that of the attraction. Once the individual decides to join the attraction, then he shifts his desired direction vector $\vec e_i$ towards the center of the attraction. For simplicity, $K_a$ and $K_0$ are set to be 1, meaning that both options are equally attractive when the individual would see nobody within his perception range. An individual $i$ is counted as an attending pedestrian if his efficiency of motion $E_{i} = (\vec{v}_{i} \cdot \vec{e}_{i})/{v_d}$ is lower than 0.05 within a range of 4~m from the center of the attraction after he decided to join there (red circles in the yellow shade area, Fig.~\ref{fig:influenceArea}). Here the individual efficiency of motion $E_i$ indicates how much the driving force contributes to the pedestrian $i$'s motion with a range from 0 to 1~\cite{Helbing_PRL2000, Kwak_PRE2013}. The average of $t_d$ is set to be 30~s.

\section{Results and Discussion}
\label{sec:3}
\begin{figure}[!t]
\includegraphics[width=\columnwidth]{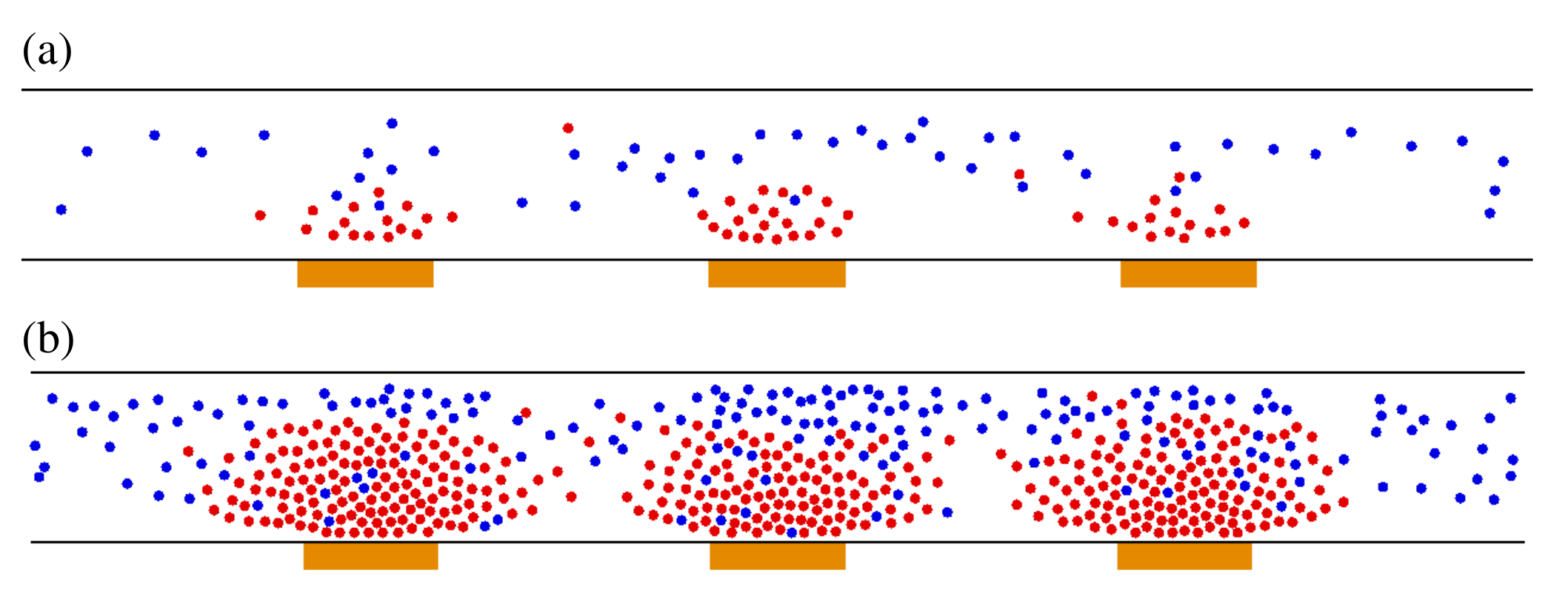}
\caption{(Color online) Snapshots of (a) a stable flow with $q$ = 0.05 and $s$ = 0.2, and (b) an unstable flow with $q$ = 0.3 and $s$ = 1.2. Red circles indicate pedestrians attracted by an attraction and blue circles for pedestrians not attracted by the attraction.}
\label{fig:snapshots} 
\end{figure}

For different levels of pedestrian influx $q$, different patterns of pedestrian movement appear. When $q$ is low, one can observe a stable flow where attendees form standstill-like clusters near the attraction and such clusters do not impede pedestrian traffic near the clusters (see Fig.~\ref{fig:snapshots}(a)). For large values of $q$, an unstable flow can be observed if the value of $s$ is large. In the unstable flow, pedestrians form swirling clusters in which pedestrians near the attractions are being pushed away from the clusters by other pedestrians. Since increasing $s$ makes more passersby head for the attractions, pedestrians tend to rush into the attraction and push others, as shown in Fig.~\ref{fig:snapshots}(b). 

In order to analyze spatial distribution of the pedestrian flow interacting with attractions, this study evaluates local quantities such as local density and local speed. Following previous studies~\cite{Helbing_PRE2007, Kwak_PRE2013, Moussaid_PNAS2011}, the local density and local speed are associated with a Gaussian distance-dependent weight function $f(d)$: 
\begin{eqnarray}\label{eq:Gauss}
f(d) = \frac{1}{\pi R^2}\exp\left(-\frac{d^2}{R^2}\right)
\end{eqnarray}

with a parameter $R = 0.7$. The local density at a location $\vec z$ and time $t$ is defined as 

\begin{eqnarray}\label{eq:LocalDensity}
\rho(\vec{z}, t) = \sum_{i}^{ }{f(d_{iz})}.
\end{eqnarray}

where $d_{iz}$ is the distance between location $\vec{z}$ and pedestrian $i$'s position. Likewise, the local speed is given as 

\begin{eqnarray}\label{eq:LocalSpeed}
V(\vec{z}, t) = \frac{\sum_{i}^{ }{\|\vec{v}_{i}\|f(d_{iz})}}{\sum_{i}^{ }{f(d_{iz})}},
\end{eqnarray}

\begin{figure}[!t]
\includegraphics[width=\columnwidth]{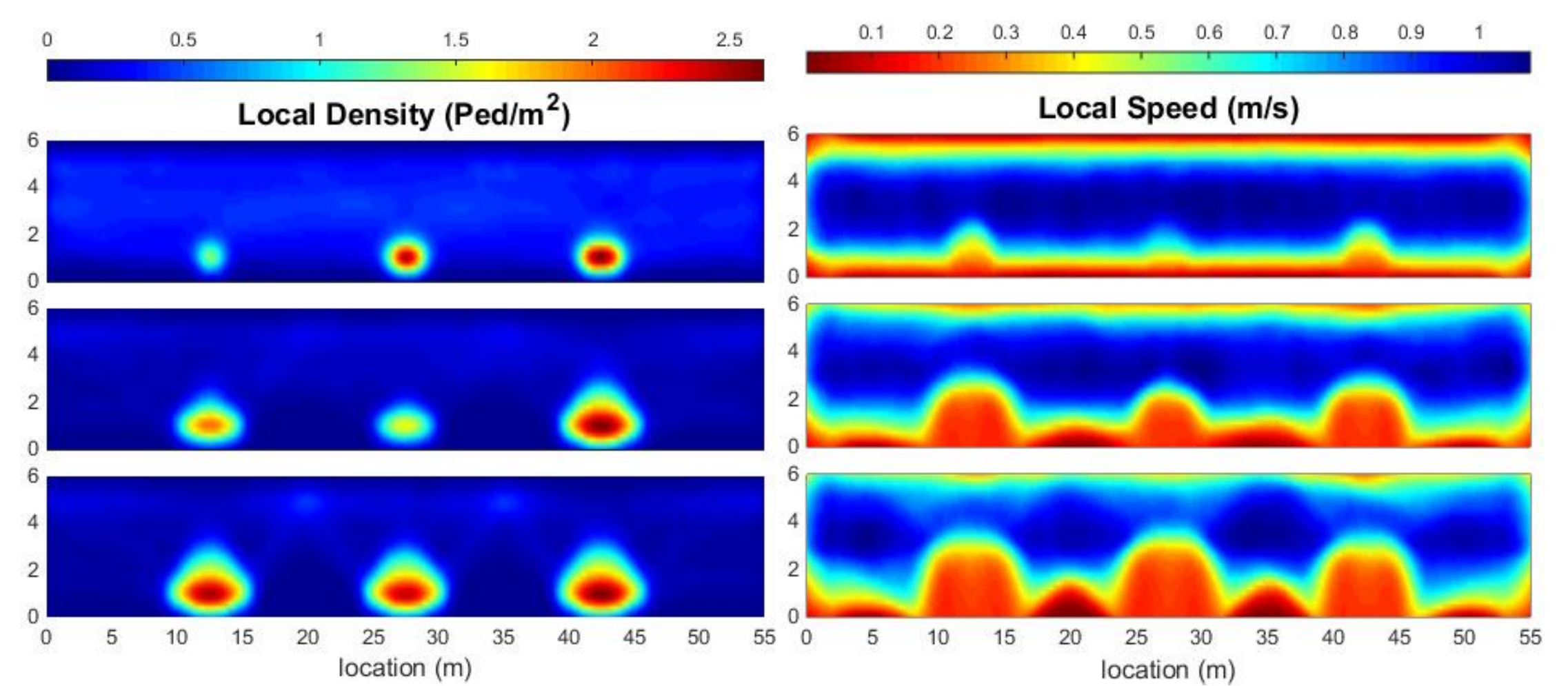}
\caption{(Color online) Local speed maps and local density maps for stable flow with a low value of $q = 0.05$~P/m/s and different values of $s$: $s = 0.2$, $0.6$, and $1.2$ (from top to bottom). In local speed maps, red and blue colors indicate lower and higher speed, respectively. In local density maps, blue and red colors indicate lower and higher density, respectively. The center of each attraction is at $x = $ 12.5, 27.5, and 42.5.}
\label{fig:heatmap1}
\end{figure}

\begin{figure}[!t]
\includegraphics[width=\columnwidth]{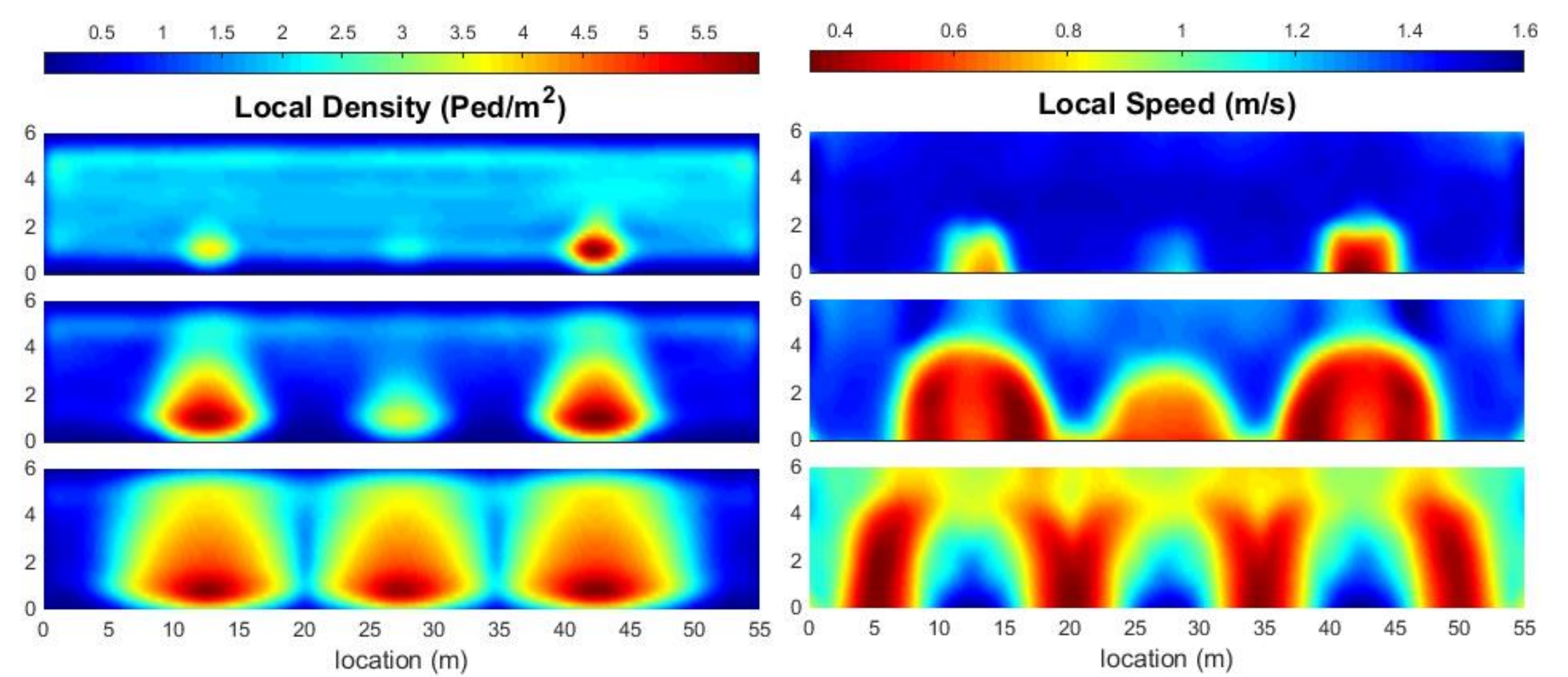}
\caption{(Color online) Local speed maps and local density maps for unstable flow with a high value of $q = 0.3$~P/m/s and different values of $s$: $s = 0.2$, $0.6$, and $1.2$ (from top to bottom).}
\label{fig:heatmap2}
\end{figure}

For different patterns, Figs.~\ref{fig:heatmap1} and \ref{fig:heatmap2} show the local speed maps $V(\vec{z},t)$ and the local density maps $\rho(\vec{z}, t)$ that have been averaged over the simulation period. In the stable flow, pedestrians form tighter clusters around the attraction as $s$ increases, resulting in higher local density and lower local speed around the attractions (see Fig.~\ref{fig:heatmap1}). In the unstable flow, one can observe higher local density as $s$ increases, similar to the observations from the stable flow. However, the local speed inside of the clusters decreases and then increases as $s$ increases while the local speed near the clusters begins to decrease when $s$ is above a certain value (see Fig.~\ref{fig:heatmap2}). 

In addition to the local quantities, one can better understand the spatial patterns of the pedestrian flow by means of collective quantities such as the collective efficiency of motion $E$ and the normalized kinetic energy $K$. As in previous studies~\cite{Helbing_PRL2000, Kwak_PRE2013}, $E$ and $K$ are measured as:

\begin{eqnarray}\label{eq:E}
E = \left\langle \frac{1}{N} \sum_{i=1}^{N} \frac{\vec{v}_{i}\cdot\vec{e}_{i}}{v_d} \right\rangle = \left\langle \frac{1}{N} \sum_{i=1}^{N} {E_i} \right\rangle
\end{eqnarray}

and

\begin{eqnarray}\label{eq:K}
K = \left\langle \frac{1}{N} \sum_{i=1}^{N} \frac{\|\vec{v}_{i}\|^2}{v_d^2} \right\rangle.
\end{eqnarray}

Here $\langle\cdot\rangle$ represents an average over the simulation period after reaching the stationary state. Similar to $E_{i}$ in the previous section, the collective efficiency reflects the contribution of the driving force in the collective pedestrian motion. The normalized kinetic energy has the value of $0$ if all pedestrians do not move, otherwise it has a positive value.

\begin{figure}[!t]
\includegraphics[width=.9\columnwidth]{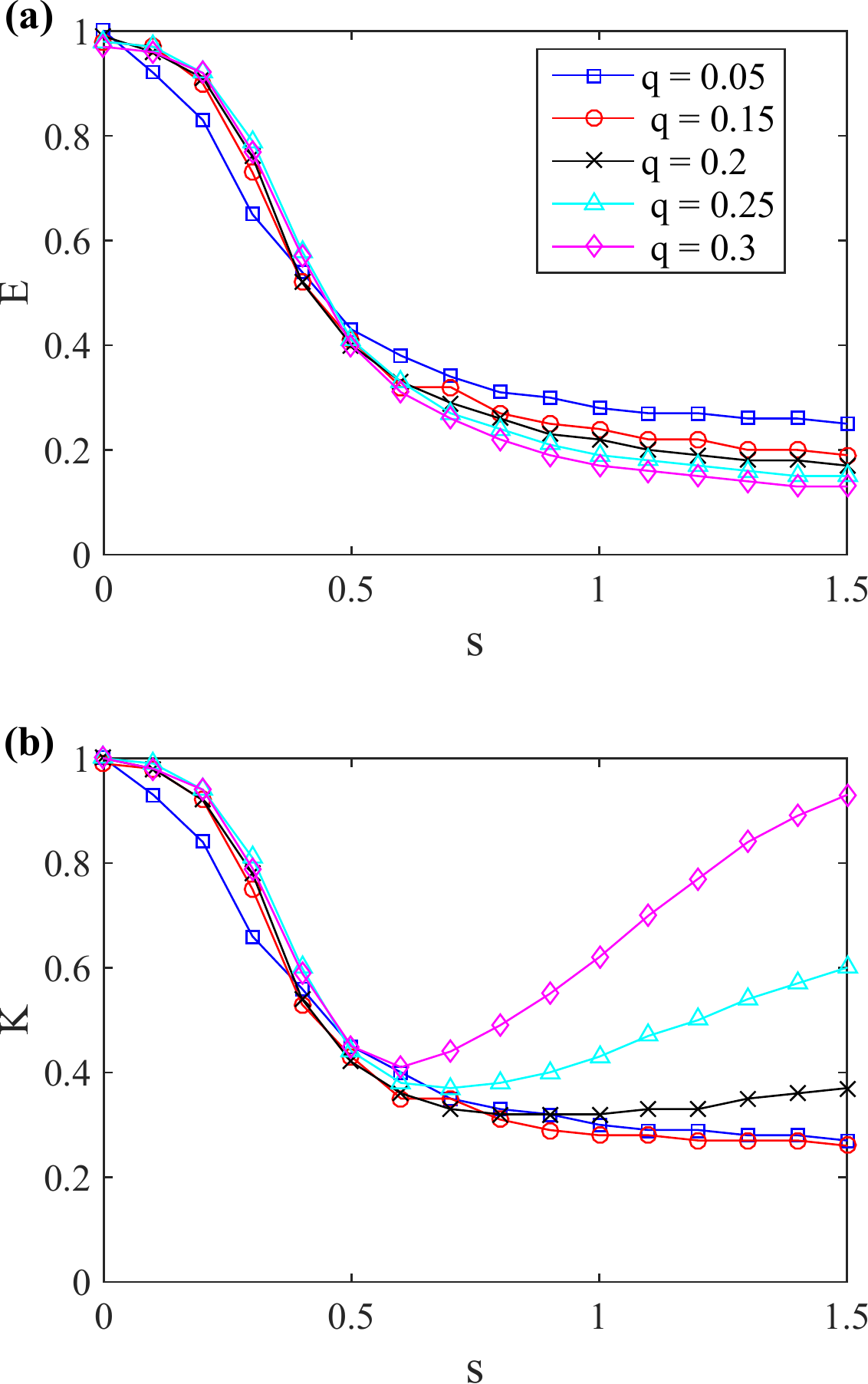}
\caption{(Color online) Numerical results of (a) the collective efficiency of motion $E(s, q)$ and (b) the normalized kinetic energy $K(s, q)$. Different symbols represent different levels of $q$. One can observe that $E(s)$ decreases as $s$ increases for each given $q$. The behavior of $K(s)$ is similar when the value of $q$ is low. However, $K(s)$ decreases and then increase against $s$ when $q$ is large.}
\label{fig:KE}
\end{figure}

Fig.~\ref{fig:KE} shows how the collective efficiency $E(s, q)$ and the kinetic energy $K(s, q)$ depend on the social influence strength $s$ and the pedestrian influx $q$. For each value of $q$, $E$ decreases as $s$ increases, indicating that more pedestrians are distracted from their initial destination due to the higher social influence (Fig.~\ref{fig:KE}(a)). Depending on $q$, $K$ reveals two distinct behaviors. First, for low values of $q$, the decreasing behavior of $K$ appears to be similar to that of $E$. This corresponds to the stable flow that can be characterized by 

\begin{equation}
\frac{\partial E}{\partial s} < 0\ \textrm{and}\ \frac{\partial K}{\partial s} < 0.
\end{equation}

Secondly, for large values of $q$, $K$ decreases and then increases as $s$ grows, indicating the unstable flow. This case can be characterized by 

\begin{equation}
\frac{\partial E}{\partial s} < 0\ \textrm{and}\ \frac{\partial K}{\partial s} > 0.
\end{equation}

\begin{figure}
\includegraphics[width=.75\columnwidth]{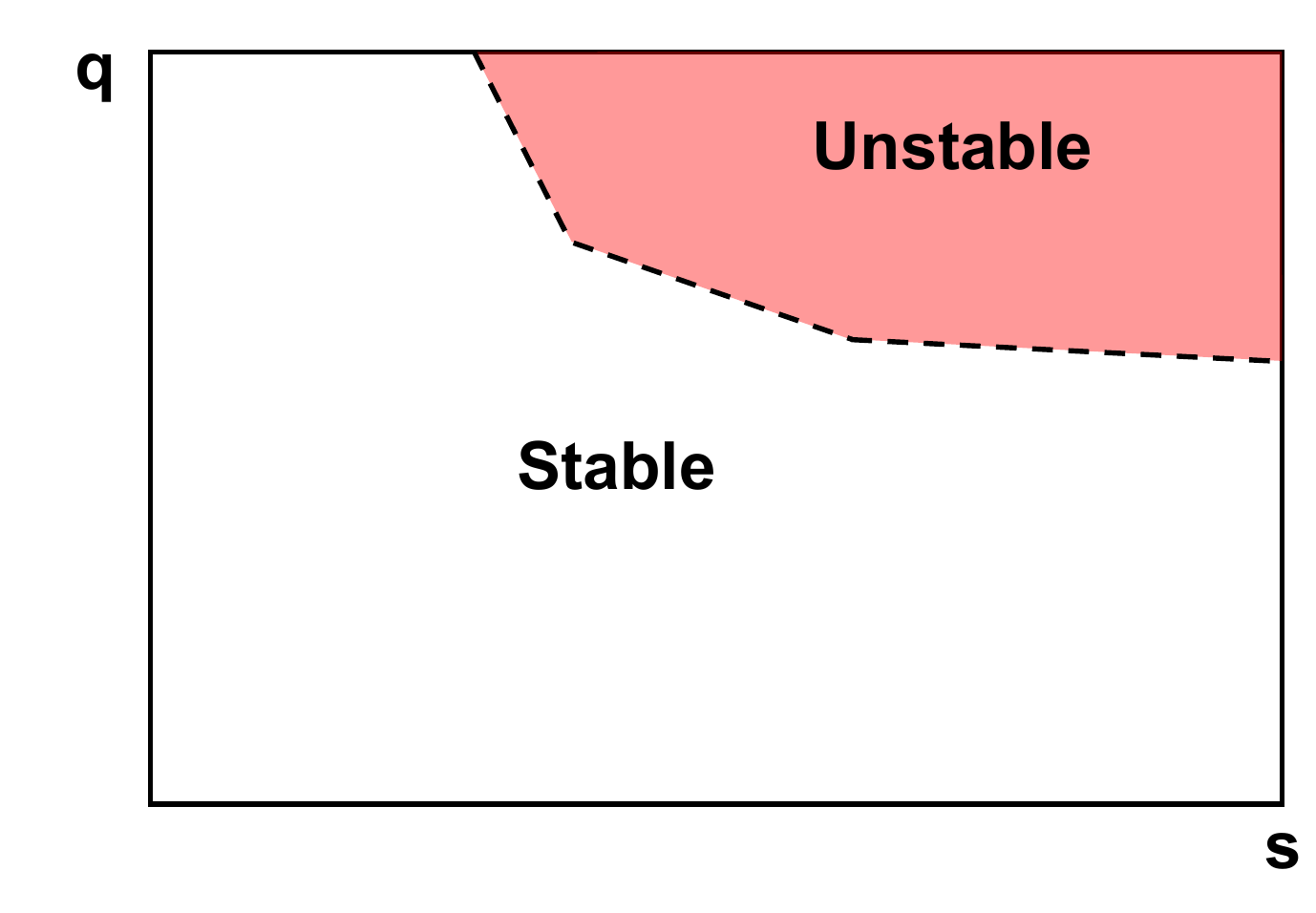}
\caption{(Color online) A schematic representation of phase diagram. The dashed line indicates ${\partial K}/{\partial s} = 0$, the boundary between the stable and unstable flow.}
\label{fig:phase}
\end{figure}

This reflects that higher $s$ yields not only more attendees around attractions but also stronger repulsion among attendees. In this case, attendees near the attraction cannot reach a standstill and they are crowded out from the cluster by other attendees because of interpersonal repulsion. Consequently, expelled attendees from the cluster impede pedestrian flow between the left and right boundaries of the corridor, and thus this parameter region can be called the unstable flow. Those different patterns of pedestrian flow can be summarized in a schematic phase diagram, as shown in Fig.~\ref{fig:phase}. 

\section{Conclusion}
\label{sec:4}
This study has numerically investigated the impact of impulse stops on pedestrian flow by employing the switching behavior model. For low pedestrian influx, one can observe a stable flow in which attendees form standstill-like clusters. When the pedestrian influx and the social influence strength are high, on the other hand, one can see an unstable flow showing crowded out attendees from the clusters. Consequently, the expelled attendees impede the pedestrian flow near the clusters. We have also provided a schematic representation of phase diagram as a summary of the study results.

We believe that our study results can provide an insight into the better management of pedestrian facilities where impulse stops may be expected to occur. For instance, during shopping holidays such as Black Friday in the United States and Singles day in China, the influx of people with extreme desire for merchandise may lead pedestrian incidents. The existence of the unstable flow suggests that controlling the pedestrian influx for expected level of social influence is necessary for safe and efficient use of pedestrian facilities.

The presented model can be further investigated for various scenarios. For instance, one can explicitly consider the capacity of the attractions, meaning that only a certain number of attendees can stay near the attractions. In addition, the length of stay $t_d$ can be associated with the number of attendees near the attractions. 

\begin{acknowledgments}
Hang-Hyun Jo gratefully acknowledges financial support by Mid-career Researcher Program through the National Research Foundation of Korea (NRF) grant funded by the Ministry of Science, ICT and Future Planning (2014030018), and Basic Science Research Program through the National Research Foundation of Korea (NRF) grant funded by the Ministry of Science, ICT and Future Planning (2014046922) and by Aalto University postdoctoral program.
\end{acknowledgments}


\end{document}